# TRANSVERSE IMPEDANCE BENCH MEASUREMENTS IN NLC/JLC ACCELERATING STRUCTURES*

N. Baboi [#], G.B. Bowden, R.M. Jones, S.G. Tantawi, J.R. Lewandowski,
SLAC, Stanford, CA 94025, USA

*Abstract*

The wire method is a more rapid and less costly method to measure impedances of RF components compared to methods using a beam. A setup using a single displaced wire to excite and measure transverse resonant modes in accelerating structures for the Next Linear Collider/Japanese Linear Collider (NLC/JLC) has been built. The RF signal is coupled into and out of the structure using two matching sections with a broadband frequency from 11 to 18 GHz. Their contribution to the scattering parameter is minimized by a calibration technique. A standing wave structure has been measured. Difficulties in accurately predicting the modal loss factors were encountered related to the approximations made and to experimental issues. The measurements are presented and comparisons with simulations are made.

## INTRODUCTION

Wakefields in accelerating structures are the main cause of transverse beam emittance growth in high energy accelerators. By appropriate design of the accelerating structures and absorption of electromagnetic energy, they must be brought below a limit where the effect on the beam is acceptable. Therefore their understanding is essential. Although modern computer codes have proved to give reliable results, measurements are important as well.

Direct measurements of wake potential simply studying their interaction with a beam. Measurements in the time domain have been made for example at the ASSET setup at SLAC on X-band accelerating structures [1], while at TTF at DESY individual dipole modes have been studied [2]. The results of these types of measurements are in good agreement with theoretical predictions [3]. Nevertheless they require long preparation times and costly beam time. Bench-top measurements are made by perturbing the field with a small bead [4] or by propagating RF through a coaxial line made of the structure and one or two thin wires [5].

In this paper we present first frequency-domain measurements using the wire method on a 11.4 GHz standing-wave structure in study for the NLC/JLC [6]. Results are discussed and comparisons with simulations are made.

## METHOD

Since their proposal in the 1970s, time domain [7] and frequency domain [8] wire measurements have been made for various microwave components. In time domain the method requires current pulses of a length shorter or comparable to the length of the bunches used in that structure. Therefore we chose to study individual modes in frequency domain. The wire method theory is rather well established for resistive wall wakefields and modes below cut-off. Several approximations exist for lumped and distributed impedances [9]. Attention must be paid particularly when these conditions are not met.

*Setup*

The measurement setup is shown in Fig. 1. The signal from a network analyzer is matched to the ports of the structure through two specially designed sections containing adapters from rectangular waveguides (WR62) to coax. These have a broadband from about 11 to 18 GHz. The wire is inserted through the adapter and can be displaced by slightly bending the thin coax tubes. We chose this solution in order to minimize discontinuities in the structure.

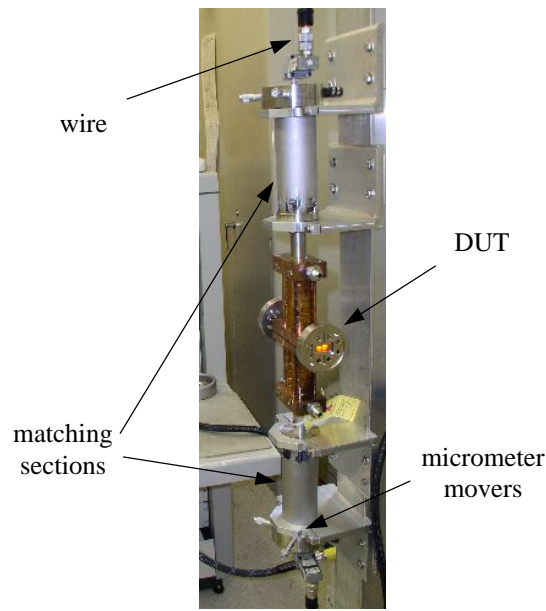

Fig. 1 Wire measurement setup

The wire is made of brass and has a diameter of 300 μm. Compared to previous measurements in single cell cavities, the wire diameter has been increased in order to provide a better contact. This has improved the stability and reproducibility of the transmission curves. For the same reason, the flanges of the matching sections have been modified to eliminate the gaskets. The reflection of the matching sections is better than –30 dB.

___________________
*Supported by the DOE, grant number DE-AC03-76SF00515
# On leave from NILPRP, P.O. Box MG-36, 76900 Bucharest, Romania

## TRL calibration

In order to obtain measurements that are hardly affected by the imperfections of the matching sections, we calibrated the measurements at the ports of the matching sections connecting to the device under test (DUT). For this we used a TRL (through-reflect-line) technique in which only a few parameters of the calibration standards need to be precisely known [10]. A delay line and an RF short were required to be built specially for this setup. The third standard is a through measurement. With this method one does not need to know precisely the reflection of the short, or the length of the delay line.

An alternative was to calibrate the network analyzer directly by modifying a calibration kit for a coaxial line to fit the parameters of our setup. Fig. 2 shows the agreement of the calibration results obtained with these two methods. The DUT is here a single cell structure.

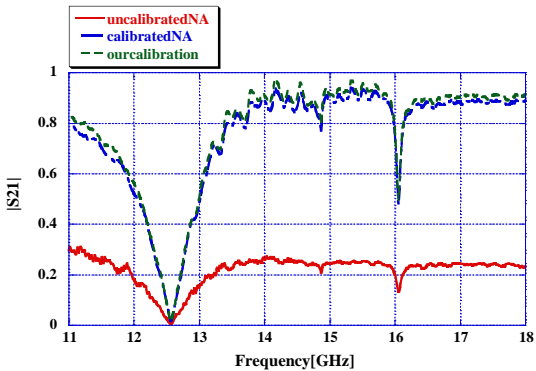

Fig. 2 TRL calibration of measurements on a single-cell structure.

## Standing-wave structure

A standing wave structure was measured in this setup (see Fig. 1). It consists of 15 cells which, prior to tuning, are identical. The structure is fed through the middle cell by a symmetrical coupler.

The main contribution to the transverse wake field is given by the first 3 dipole passbands [11]. The setup allows us to measure the first two bands plus the fundamental monopole band.

## RESULTS

The two ends of the wire were approximately centered. Then the wire was displaced using two transverse micrometers with respect to this position. Moving the wire rather than measuring at a fixed position allows us to reduce statistical errors. Moreover, one can fit data and ascertain the loss factor in the limit of centered wire [12]. At the same time the alignment of the wire can be based on the dipole signal.

## Measurements with various wire offsets

Fig. 3 shows the transmission parameter $S_{21}$ measured on the standing-wave structure for various transverse positions of the wire. One can distinguish the fundamental monopole passband between about 11.8 and 12.5 GHz. It remains essentially unchanged with changing wire offset.

All modes at higher frequencies are from the first and second dipole passbands in interaction with the TEM coaxial mode. The amplitudes of the peaks change with the wire position. One can see that they are smallest for a wire offset around 0.5 mm. The frequency of the modes also shifts as the displacement of the wire changes. In the limit of zero wire offset, the frequency goes to the frequency of the unperturbed dipole.

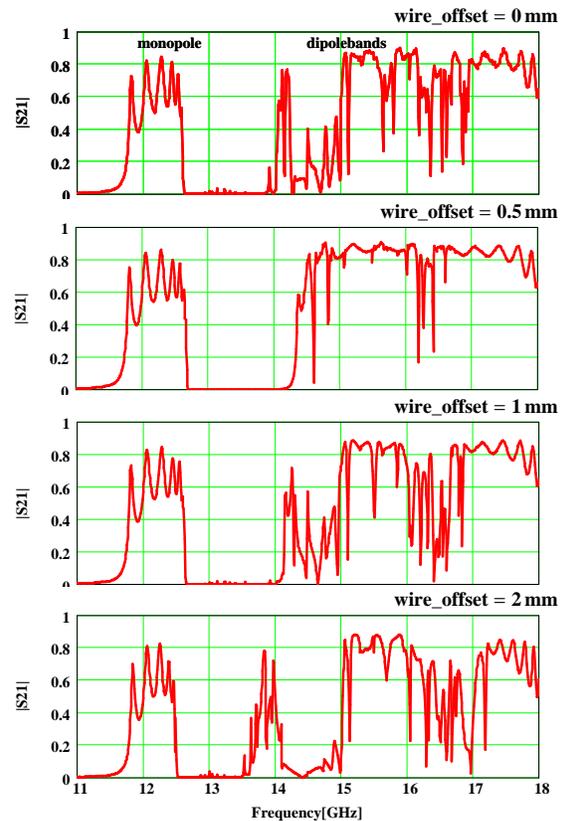

Fig. 3 $S_{21}$ for various transverse positions of the wire.

## Comparison to simulations

The properties of the standing wave structure have been simulated with the HFSS code (High Frequency Structure Simulator). The S-parameters of a single cell and of the end cell plus the beampipe were obtained. A wire with 300 mm in diameter was placed 1 mm off-center. Using a cascading technique, the result for a 15-cell structure was obtained [13].

Fig. 4 compares the simulation with the measurement made for a relative wire displacement of 0 mm which seems to match it best. The agreement for the monopole band is good apart from a frequency shift. This is due to neglecting the input coupler in the simulations. Also the actual structure was tuned to obtain a flat accelerating field.

The dipole bands are more difficult to compare. They depend strongly on the alignment of the wire and clearly the two curves are not made at the same wire offset. Also it is likely that the wire was tilted with respect to the structure axis and therefore some modes have higher amplitude in the simulation and others in the measurement. The alignment of the structure cells may also play a role.

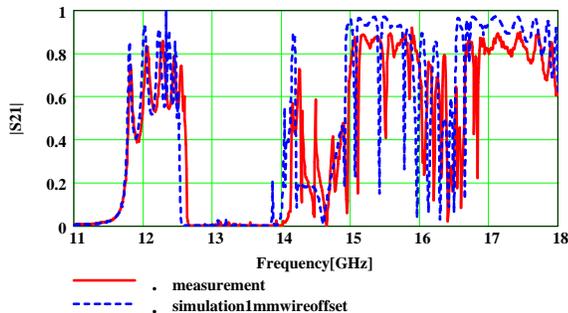

Fig. 4 Comparison of measurement with the wire placed at relative position of 0mm with simulation for a wire offset of 1mm.

The coupling impedance and the loss factors of the structure are related to the S-parameters. The correlation is nevertheless not straightforward for accelerating structures. The theory is rather well established for lumped impedances, and for distributed impedances, below cut-off. Further studies are in progress in this sense.

## SUMMARY

Initial wire measurements of a NLC/JLC accelerating structure have been undertaken. Special matching sections have been designed and built in order to allow for a relatively easy mounting and displacement of a thin brass wire. A TRL calibration technique gives good results when applied directly on the network analyzer, as well as post-measurement. By moving the wire, excitation of dipole modes was observed. The simulations show good agreement with the measured monopole band, and give a qualitative comparison for the dipole. Nevertheless better wire tilt alignment is needed. Further investigations on inferring the coupling impedance of the dipole bands is underway.

## ACKNOWLEDGEMENTS

We thank our colleagues from the Accelerator Structures and High Power Microwave groups for the useful discussions.